\documentstyle{article}
\newcommand{\beq}{\begin{equation}}
\newcommand{\eeq}{\end{equation}}
\newcommand{\beqa}{\begin{eqnarray}}
\newcommand{\eeqa}{\end{eqnarray}}

\def\ra{\rightarrow}
\def\x{\times}
\def\etal{\it etal,}

\newcommand{\RMP}[3]{{\elevenit Rev. Mod. Phys.} {\elevenbf #1}, #2 (19#3)}
\newcommand{\PR}[3]{{\elevenit Phys. Rev.} {\elevenbf #1}, #2 (19#3)}
\newcommand{\PL}[3]{{\elevenit Phys. Lett.} {\elevenbf #1}, #2 (19#3)}

\newcommand{\PRL}[3]{{\elevenit Phys. Rev. Lett.} {\elevenbf #1}, #2 (19#3)}
\newcommand{\NP}[3]{{\elevenit Nucl. Phys.} {\elevenbf #1}, #2 (19#3)}

\def\ev{\; {\rm eV} }

\def\etal{ {\it et al}.}
\font\tenbf=cmbx10
\font\tenrm=cmr10
\font\tenit=cmti10
\font\elevenbf=cmbx10 scaled\magstep 1
\font\elevenrm=cmr10 scaled\magstep 1
\font\elevenit=cmti10 scaled\magstep 1

\renewenvironment{thebibliography}[1]
 { \elevenrm
   \begin{list}{\arabic{enumi}.}
    {\usecounter{enumi} \setlength{\parsep}{0pt}
     \setlength{\itemsep}{3pt} \settowidth{\labelwidth}{#1.}
     \sloppy
    }}{\end{list}}

\begin{document}
\begin{center}{{\tenbf IMPLICATIONS OF SOLAR AND ATMOSPHERIC NEUTRINOS\\}
\vglue 1.0cm
{\tenrm PAUL LANGACKER\\}
\baselineskip=13pt
{\tenit University of Pennsylvania \\  Philadelphia, Pennsylvania, USA
19104-6396\\} \vglue 0.8cm
{\tenrm ABSTRACT}}
\end{center}
\vglue 0.3cm
{\rightskip=3pc
 \leftskip=3pc
 \tenrm\baselineskip=12pt
 \noindent
The implications of the deficit of solar neutrinos are discussed.  If all of
the
experiments are taken literally the relative suppressions render
an astrophysical
explanation unlikely.  Allowing MSW conversions, the data simultaneously
determine the temperature of the core of the sun to within five
percent.
The implications of the atmospheric $\nu_\mu/\nu_e$ ratio are
briefly discussed.
\vglue 0.6cm}
{\elevenbf\noindent 1. Solar Neutrinos and Cool Sun Models}
\vglue 0.2cm
\baselineskip=14pt
\elevenrm
The predictions of two recent theoretical studies are shown in
Table~\ref{tab1}.  There is reasonable agreement between them,
especially for the gallium experiments.  However, the
Bahcall-Pinsonneault (PB)~\cite{bp} calculation predicts a somewhat
higher $\,^8B$ flux than that of Turck-Chi\`{e}ze (TC) \cite{tc}.
These are compared with the experimental
results~\cite{homestake}-\cite{sage}
in Table~\ref{tab2}.
The standard solar model is
not in agreement with the data for any reasonable range of the
uncertainties~\cite{BB}, and is therefore excluded.

Still possible is some nonstandard solar model (NSSM), which
may differ from the SSM by  new physics inputs such as weakly
interacting massive particles (WIMPs), a large core magnetic field,
core rotation, {\it etc.} \ Most of these models affect the solar
neutrinos by leading to a lower core temperature,
{\it i.e.}, $T_c<1$, where $T_c = 1 \pm 0.006$
corresponds to the SSM. All reasonable models lead to a
larger suppression of the Kamiokande rate (which is
essentially all $\,^8B$) than that of Homestake (which has
in addition a nontrivial component of $\,^7Be$ neutrinos), in
contrast to the data.

Following Bahcall and Ulrich~\cite{bu} the temperature dependence  is $\varphi
(\,^8B) \sim T^{18}_c$, $\varphi(\,^7Be) \sim T_c^8$. We assume that the $pp$
flux is reduced by a factor $f(pp)$  chosen so that the total solar
luminosity remains constant.
The expected counting rates $R$ for
each experiment relative to the expectations of the standard solar
model are then~\cite{BHKL}
\beqa R_{c\ell} = &0.26 \pm 0.04 = & ( 1 \pm 0.033)  [ 0.775 (1
\pm 0.10) T_c^{18}
 + 0.150 ( 1 \pm 0.036) T_c^8 + \;{\rm small}\;] \nonumber \\
R_{\rm Kam} = & 0.50 \pm 0.07 = & (1 \pm 0.10) T_c^{18} \nonumber \\
R_{\rm Ga} = & 0.54 \pm 0.11 = & ( 1 \pm 0.04) [ 0.538 ( 1 \pm
0.0022)
f(pp) \nonumber \\  && + 0.271 (1 \pm 0.036) T_c^8
 + 0.105 (1 \pm 0.10) T_c^{18} + \; {\rm small}\, ]. \label{eqa} \eeqa
The overall uncertainties  are from the nuclear detection
cross-sections, and those which multiply the individual
flux components are from the relevant
reactions in the sun,  correlated from
experiment to experiment.

The best fit  is for $T_c = 0.92 \pm 0.01$,
an enormous deviation from the SSM.  Even worse, it
is a terrible fit: $\chi^2 = 20.6$ for $2\; d.f.$, which is
statistically excluded at the $99.9\% \,cl$.  If we accept the
experimental values, a cool sun model  cannot account for the
data~\cite{BHKL}.
This conclusion is  more general than the specific exponents
assumed.

\begin{table} \centering
\begin{tabular}{|c|c|c|} \hline
Theory & SSM (BP) & SSM (TC) \\ \hline
Homestake (Cl) & $8 \pm 1 $ SNU & $6.4 \pm 1.3$ SNU \\
Kamiokande & $1 \pm 0.14$ (arb units) & $0.77 \pm 0.20$ \\
gallium & $132 \pm 7 $ SNU & $125 \pm 5 $ SNU \\ \hline
\end{tabular}
\caption{Predictions of Bahcall-Pinsonneault~(BP)~\protect\cite{bp}
and Turck-Chi\`{e}ze~(TC)~\protect\cite{tc} for the solar neutrino
fluxes.  All uncertainties are at one standard deviation.}
\label{tab1}
\end{table}

\begin{table} \centering
\begin{tabular}{|c|c|c|c|} \hline
 & Rate & Rate/SSM (BP) & Rate/SSM (TC) \\ \hline
Homestake & $2.1 \pm 0.3$ SNU & $0.26 \pm 0.04$ & $0.33 \pm
0.05$\\
Kam-II (1040 days) &  & $0.47 \pm 0.05 \pm 0.06 $   & \\
Kam-III (395 days) &  & $0.56 \pm 0.07 \pm 0.06 $  & \\
Kam-II + III      &  & $0.50 \pm 0.07 $ & $0.65 \pm 0.09$ \\
\raisebox{.9ex}{(prelim syst.)}&  & & \\
GALLEX & $83 \pm 19 \pm 8 $ SNU & $0.63 \pm 0.14$ & $0.67 \pm
0.15 $ \\
SAGE (90 + 91) & $58^{+17}_{-24} \pm 14$ SNU  & $0.44 \pm 0.19$ &
$0.47 \pm 0.20$ \\
GALLEX + SAGE & $71 \pm 15$ SNU & $0.54 \pm 0.11$ & $0.57 \pm
0.12$ \\
\hline
\end{tabular}
\caption{The observed rates, and the rates relative to the  calculations of BP
and TC.} \label{tab2}
\end{table}

\vglue 0.6cm
{\elevenbf\noindent 2. MSW Conversions}
\vglue 0.2cm
There have been a number of recent studies of the MSW
solution~\cite{BHKL}-\cite{18a}.  There are two solutions
for oscillations into active neutrinos $(\nu_\mu$ or $\nu_\tau)$,
the non-adiabatic (small
mixing angle) and the large-angle.
The non-adiabatic solution gives a much better fit~\cite{BHKL}.  In this
region there is more suppression of the intermediate energy $\,^7Be$
neutrinos, accounting for the larger suppression seen by Homestake.
The large-angle fit is much poorer, corresponding to
$\chi^2= 3.8$ for $1 \;df$, because there the survival
probability varies slowly with neutrino energy.  One
can also consider the possibility that the $\nu_e$ is oscillating into
a sterile neutrino. There is no large-angle
solution\footnote{ The large angle solution for sterile neutrinos is
also most likely excluded by nucleosynthesis arguments,
while  non-adiabatic parameters are allowed.} at 90\% C.L.  There is a
non-adiabatic solution but even that yields a relatively poor fit $\chi^2 =
3.6$
for 1~$df$.

It is also interesting to consider MSW
oscillations for an arbitrary core temperature $T_c$,
that is for NSSM. One now has three parameters, $T_c$,
$\sin^22\theta$, and $\Delta m^2$.  There are
sufficient constraints to determine all three \cite{BHKL}.  There is
an expanded non-adiabatic solution with $T_c
= 1.02^{+0.03}_{-0.05}$ at 90\%~C.L.  Similarly, there is a
large-angle solution with $T_c = 1.04^{+0.03}_{-0.04}$.  Thus the core
temperature is measured by the solar neutrino
experiments,\footnote{Of course, one of the original motivations for
the solar neutrino experiments was to probe the core of the sun.} even
allowing for the complication of MSW oscillations.  It
is consistent with the standard solar model prediction $T_c = 1 \pm
0.0057$.

\vglue 0.6cm
{\elevenbf\noindent 3. Atmospheric Neutrinos}
\vglue 0.2cm
The predicted fluxes ${\nu_\mu}$ and ${\nu_e}$ produced by the
interactions of cosmic rays in the atmosphere are uncertain by around
20\%.  However, the ratio ${\nu_\mu}/{\nu_e}$ is believed to be
accurate to $\sim$~5\% \cite{penn}.  There are additional
uncertainties associated with interaction cross-sections, particle
identification, {\it etc}. The Kamiokande and IMB groups have observed
a deficit in the ratio of contained muon and electron events
\beq \frac{(\mu/e) |_{\rm data} }{ (\mu/e) |_{\rm
theory} } = \left\{ \begin{array}{ccc} 0.65 \pm 0.08 \pm 0.06 & ,
& {\rm Kamiokande\ \protect\cite{8a}} \\ 0.54 \pm 0.05 \pm 0.12 & . &
{\rm IMB\ \protect\cite{imb}} \end{array} \right. \eeq
This effect, if real, suggests the possibility of $\nu_\mu \ra
\nu_\tau$ or possibly $\nu_\mu \ra \nu_e$.  It probably is not
compatible with sterile neutrino oscillations $\nu_\mu \ra \nu_s$,
because for the relevant parameter range the extra sterile neutrino would
violate the nucleosynthesis bound.
The oscillation
hypothesis requires a mass range $\Delta
m^2 \sim (10^{-3} {-1}) \ev^2$, larger than that relevant to the solar
neutrinos, and large mixing angles such as $\sin^22 \theta \sim 0.5$.

\begin{figure}
\vspace{9.5cm}
\caption{Allowed regions for MSW conversions of $\nu_e \ra \nu_\mu$ or
$\nu_\tau$, from \protect\cite{BHKL}.  The 90\% c.l. $(\Delta \chi^2 =
4.6)$ regions allowed by the Homestake, Kamiokande, and gallium
experiments and by the combined fit are shown.  The astrophysical and
nuclear uncertainties are included.}
\label{fig2}
\end{figure}

\vglue 0.6cm
{\elevenbf\noindent 4. Implications}
\vglue 0.2cm
The $\Delta m^2$ range suggested by the solar
neutrinos is compatible with the general range expected in
quadratic up-type seesaw models, such as in grand unified theories
\cite{BHKL}, for which $m_{\nu_i} \sim m_{u_i}^2/M_N$,
where $u_i = u,c,t$ and $M_N$ is the heavy neutrino mass.  For $M_N
\sim 10^{11} - 10^{16}$~GeV one obtains the appropriate mass range,
for oscillations into $\nu_\tau$ for $M_N\sim 10^{16}$ GeV, and into
$\nu_\mu$ for $M_N \sim 10^{11}$~GeV.  However, the simplest models
predict equal lepton and quark mixing angles, $V_{\rm lepton} = V_{\rm
CKM}$, which is not satisfied by the data unless $T_c$ is far from the
SSM \cite{BHKL}.

The various hints suggest two general scenarios.  One could
have $\nu_e \ra \nu_\mu$ in the sun for $m_{\nu_e} \ll m_{\nu_\mu}
\sim 3 \x 10^{-3}$~eV, with $\nu_\tau$ a component of the dark matter
($m_{\nu_\tau} \sim few $~eV).  This pattern is compatible with the
mass predictions of GUT-type seesaws.  However, in this scenario there
is no room for oscillations to account for the atmospheric neutrinos.
A separate possibility is that again $\nu_e \ra \nu_\mu$ in the sun,
with $\nu_\mu \ra \nu_\tau$ oscillations with $m_{\nu_\tau} \sim (0.1 -
0.6)$~eV for the atmospheric neutrinos.  In this case there
would be no room for hot dark matter.  This second solution requires
large leptonic mixings, $\sin^2 2 \theta_{\nu_\mu \nu_\tau} \sim 0.5$.

\end{document}